\documentstyle[aps,prl,multicol,floats,epsf]{revtex}

\begin{document}

\draft
\preprint{}
\wideabs{

\title{Vortex phase transformations probed by the local ac response of 
Bi$_{2}$Sr$_{2}$CaCu$_2$O$_{8+\delta}$ single crystals with various doping}

\author{Yoichi Ando$^1$ and K. Nakamura$^{1,2}$}

\address{$^1$Central Research Institute of Electric Power
Industry, Komae, Tokyo 201-8511, Japan}
\address{$^2$Department of Energy Science, Tokyo Institute of Technology,
Nagatsuta, Yokohama 226-8502, Japan}

\date{\today}

\maketitle

\begin{abstract}
The linear ac response of the vortex system is measured locally in 
Bi$_{2}$Sr$_{2}$CaCu$_2$O$_{8+\delta}$ crystals at various doping,
using a miniature two-coil mutual-inductance technique.
It was found that a step-like change in the local ac response takes
place exactly at the first-order transition (FOT) temperature $T_{FOT}(H)$
determined by a global dc magnetization measurement.
The $T_{FOT}(H)$ line in the $H$-$T$ phase diagram becomes steeper 
with increasing doping.
In the higher-field region where the FOT is not observed, the local
ac response still shows a broadened but distinct feature, 
which can be interpreted
to mark the growth of a short-range order in the vortex system.
\end{abstract}

}
\narrowtext

The vortex phase diagram of the highly-anisotropic high-$T_c$
superconductor Bi$_{2}$Sr$_{2}$CaCu$_2$O$_{8+\delta}$ (BSCCO)
has been intensively studied in the past few years.
In superconductors, strong supercurrents flow near the surface, which
produce nonuniform magnetic-field distribution in the sample.
Such nonuniformity broadens the thermodynamic phase transitions 
and thereby hinders the study of the phase diagram.
Also, the surface (or edge) currents produce a geometrical barrier 
\cite{geometrical} in flat samples at low fields, which
gives rise to a hysteretic behavior \cite{Majer}.
In higher fields, the Bean-Livingston surface barrier is known
to be strong in BSCCO and this makes the global properties of the 
vortex system complicated \cite{Fuchs-local_ac}.
A useful way to get rid of the effects of the surface 
and the magnetic-field nonuniformity is to measure the 
electromagnetic properties locally.
There have been a number of efforts along this line
\cite{Majer,Fuchs-local_ac,Zeldov,Tamegai,Ando,Doyle}, and
the true nature of the vortex phases of BSCCO is beginning
to be fully understood.
For example, local magnetization measurements using microscopic
Hall probes have found, quite conclusively, the presence of a 
first-order transition (FOT) of the vortex system \cite{Zeldov}.
With the improvement of instrumentation and crystal quality,
it has become clear that the first-order transition can also be
determined as a step in the global dc magnetization
measured with SQUID magnetometer \cite{Pastoriza,Hanaguri,Watauchi}.

The miniature two-coil mutual-inductance technique \cite{Jeanneret}
has been used
for the study of the vortex phase diagram of BSCCO 
\cite{Ando,Doyle}.
With this technique, a small ac perturbation field is applied 
near the center 
of the crystal and therefore the the surface barrier, which hinders 
vortex entry and exit at the edge, has minimal effect on the
measured response. 
Because of this advantage, a sharp distinct change in the local ac 
response has been observed \cite{Ando,Doyle} 
and such a feature has been associated 
with a decoupling transition \cite{Glazman,Daemen,Ikeda}
of the vortex lines.
It is naturally expected that the \lq \lq decoupling line"
thus determined is identical to the FOT measured by dc magnetization
measurements, although there has been no direct comparison between the
two phenomenon measured on an identical sample.
Since it is known that the first-order transition in the dc
magnetization has a critical point and thus disappears above a
certain field \cite{Zeldov}, 
it is intriguing how the \lq \lq decoupling" signal
of the miniature two-coil technique transforms at higher fields, 
above the critical point. 
In fact, the nature of the vortex matter in the field range above 
the critical point is still controversial 
\cite{Fuchs-Tx,Forgan,Horovitz}; 
since the ac technique can probe
the growth of the correlation lengths 
of the vortex system \cite{Ando-YBCO},
it is expected that the local ac measurement using the two-coil
technique gives a new insight into the vortex phase transformations.

In this paper, we present the results of our miniature two-coil 
measurements and the global dc magnetization measurements 
on the same crystals.  It is found that these two techniques
detect the anomaly at the same temperature $T_{FOT}(H)$, 
directly demonstrating 
that the two phenomena are of the same origin.  
We measured crystals with three different dopings and confirmed that
the result is reproducible among systems with different anisotropy.
In higher fields where the FOT is not observed by the global 
dc magnetization measurement, a distinct feature is still observable
in the local ac response and
the position of such feature is weakly frequency dependent.
We discuss that the frequency-dependent feature above the critical point
is likely to originate from the growth of a short-range order 
in the vortex system.

The single crystals of BSCCO are grown with a floating-zone method and
are carefully annealed and quenched to obtain uniform oxygen content
inside the sample.  We obtained three different dopings by annealing
the crystals at different temperatures in air; annealing at
800$^{\circ}$C for 72 hours gives 
an optimally-doped sample with $T_c$=91 K (sample A), 
650$^{\circ}$C for 100 hours gives a lightly-overdoped 
sample with $T_c$=88 K (sample B), and 400$^{\circ}$C for 10 days 
gives an overdoped sample with $T_c$=80 K (sample C).
All the samples have the transition width of less than 1.5 K.
A tactful quenching at the end of the anneal is essential for
obtaining such a narrow transition width.
$T_c$ is defined by the onset temperature of the Meissner signal
in the dc magnetization measurement.
The crystals are cut into platelets with lateral sizes larger than 
3 $\times$ 3 mm$^2$ and the thickness of the samples are typically
0.02 mm.  We used a very small (0.6 mm diameter) coaxial set of pickup 
and drive coils for our two-coil mutual-inductance measurements
(see the inset to Fig. 1).
The details of our technique have been described elsewhere 
\cite{Ando,Ando-YBCO}.
The amplitude of the drive current $I_d$ was 7.5, 7.5, and 1.0 mA for 
the measurements of samples A, B, and C, respectively.  
The linearity of the measured voltage
with respect to $I_d$ was always confirmed.
These $I_d$ produce the ac magnetic field of about 0.01 - 
0.1 G at the sample.
We emphasize that our two-coil geometry mainly induces and detects 
shielding currents flowing near the {\it center} of the sample, 
while usual
ac-susceptibility measurements are most sensitive to shielding currents
flowing near the {\it edge} of the sample.
All the two-coil measurements are done in the field-cooled procedure.
The global dc magnetization measurements are done with a Quantum Design
SQUID magnetometer equipped with a slow temperature-sweep
operation mode.

\begin{figure}[t]
\epsfxsize=7.5cm
\centerline{\epsffile{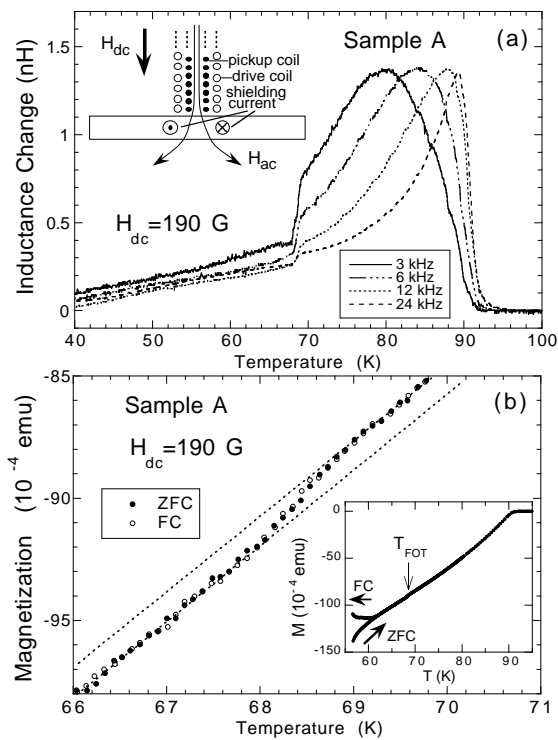}}
\vspace{2mm}
\caption{(a) $T$ dependence of the in-phase signal from 
the two-coil measurement on sample A in 190 G, taken at
3, 6, 12, and 24 kHz.
Inset shows a schematic of our two-coil measurement.
(b) $T$ dependence of the dc magnetization of sample A in 190 G.
Both the zero-field-cooled (ZFC) data and the field-cooled (FC) data
are shown. The dotted lines are guides to the eyes.
Inset show the dc magnetization in a wider temperature range.}
\label{fig:1}
\end{figure}

Figure 1(a) shows the temperature dependence of the in-phase signals of 
our two-coil measurement on sample A in 190 G, taken at
various frequencies from 3 kHz to 24 kHz.
To compare the signals from different frequencies, the data are
plotted in the unit of inductance change.
It is apparent that there is a frequency-independent step-like change 
at a temperature $T_d$, which is 68.5 K here.
The temperature dependence of the global dc magnetization in the
same field is shown in Fig. 1(b), which shows that the FOT is
taking place at exactly the same temperature as the step-like change 
in the two-coil signal.

According to the linear ac-response theory of the vortex system,
the ac response is governed by the ac penetration depth $\lambda_{ac}$
\cite{Brandt,Coffey}.
$\lambda_{ac}$ in our configuration is related to the in-plane 
resistivity $\rho_{ab}$ in the manner 
$\rho_{ab}$=Re$(i\omega \mu_0 \lambda_{ac}^2)$ \cite{Ando-YBCO}.
It has been reported that the apparent resistivity 
measured in the mixed state
of BSCCO is largely dominated by the surface current 
\cite{Fuchs-local_ac}.
Recent measurement of the bulk and surface contributions to the
resistivity found \cite{Fuchs-bulk/strip} 
that the bulk contribution shows a sharp change
at the FOT, while the surface contribution is governed by the surface
barrier and shows a broader change.
Since our measurement is not sensitive to the
edge current, it is expected that $\lambda_{ac}$ of our measurement 
reflects mainly the bulk resistivity.  
Therefore, the step-like change in the local ac response
is most likely to originate from the reported sharp change in the bulk
resistivity \cite{Fuchs-bulk/strip}.  
We note that there has been a confusion about the
origin of the step-like change in the local ac response measured by the
miniature two-coil technique and it was discussed that the source of the
sudden change may be related to a change in the $c$-axis resistivity
\cite{Ando,Doyle}.

\begin{figure}[t]
\epsfxsize=6.5cm
\centerline{\epsffile{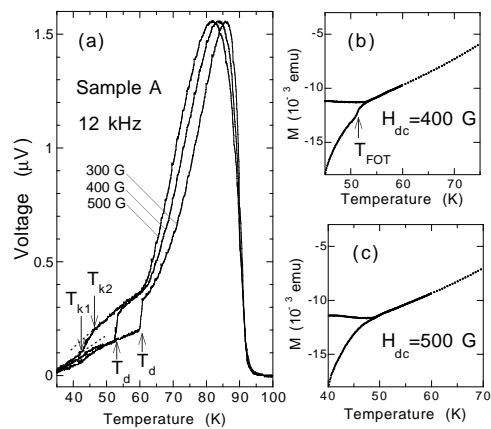}}
\vspace{2mm}
\caption{(a) $T$ dependence of the in-phase signal of the
two-coil measurement on sample A in 300, 400, and 500 G.
The dotted lines near $T_{k1}$ and $T_{k2}$ are guides to the eyes.
(b) and (c): $T$ dependence of the dc magnetization 
in 400 and 500 G, respectively.}
\label{fig:2}
\end{figure}

Figure 2(a) shows the $T$ dependence of the in-phase signals of our
two-coil measurement on sample A in three different magnetic fields.
We observed that the sharp step-like change in the two-coil signal 
becomes broadened when the magnetic field exceeds a certain limit 
$H_{lim}$;
in the case of sample A, the step-like change is observed in up to 400 G,
but becomes broadened at 500 G.
It was found that this $H_{lim}$ corresponds to the magnetic-field value
at the critical point of the FOT; namely, the FOT in the
dc magnetization measurement also disappears in fields above $H_{lim}$.
Figures 2(b) and 2(c) show that 
the FOT is observed in the dc magnetization
at 400 G but is not detectable at 500 G.
This is also a clear evidence that the origin of the step-like change 
in the two-coil signal is the FOT.

In Fig. 2(a), the 500-G data do not show a step-like change, but 
clear changes in the slope at two separate temperatures, 
$T_{k1}$ and $T_{k2}$, are discernible.  
The signal changes much rapidly between $T_{k1}$ and $T_{k2}$ compared
to the temperatures outside of this region, so the data look like that
the step-like change at $T_{d}$ is broadened to the temperature
region of $T_{k1}<$$T$$<T_{k2}$.
Figure 3 shows the in-phase signals of sample A in 600 G, 
which is above $H_{lim}$, taken at various frequencies.
Apparently, $T_{k1}$ and $T_{k2}$ inferred from the 600-G data change 
with frequency, although the change is small.
This indicates that $T_{k1}$ and $T_{k2}$ do not mark a true
phase transition but mark a crossover.

\begin{figure}[t]
\epsfxsize=7.0cm
\centerline{\epsffile{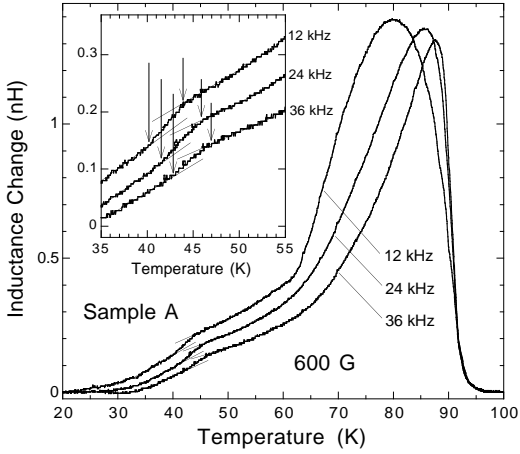}}
\vspace{2mm}
\caption{$T$ dependence of the in-phase signal of 
the two-coil measurement on sample A in 600 G taken at 12, 24, and 36 kHz.
Inset shows a magnified view of the data near $T_{k1}$ and $T_{k2}$, which
are marked by arrows.  The thin solid lines are guides to the eyes.}
\label{fig:3}
\end{figure}

Figures 4(a) and 4(b) show the in- and out-of-phase signals of samples 
B and C, respectively, in two selected magnetic fields below and
above $H_{lim}$.  Also in these two samples, the $T$ dependence of the 
two-coil signals show a step-like change in magnetic fields 
below $H_{lim}$, while the change is broadened in $H$$>$$H_{lim}$. 
Figure 5 shows the $T_{d}(H)$ lines for the three samples determined
by our two-coil measurements.
Clearly, the $T_{d}(H)$ line tends to be steeper for more overdoped
samples.
The $T_{FOT}$ data obtained from the dc magnetization are also plotted
in Fig 5; apparently, $T_{d}(H)$ and $T_{FOT}(H)$ agree very well in all
the three samples. 
The inset to Fig. 5 shows the $T_{d}(H)$ lines together with the
$T_{k1}(H)$ and $T_{k2}(H)$ lines at higher fields
(determined with 12 kHz), plotted versus
normalized temperature $T/T_c$.
The $T_{k1}(H)$ and $T_{k2}(H)$ lines are much steeper compared to the
$T_{d}(H)$ line.

\begin{figure}[t]
\epsfxsize=6.0cm
\centerline{\epsffile{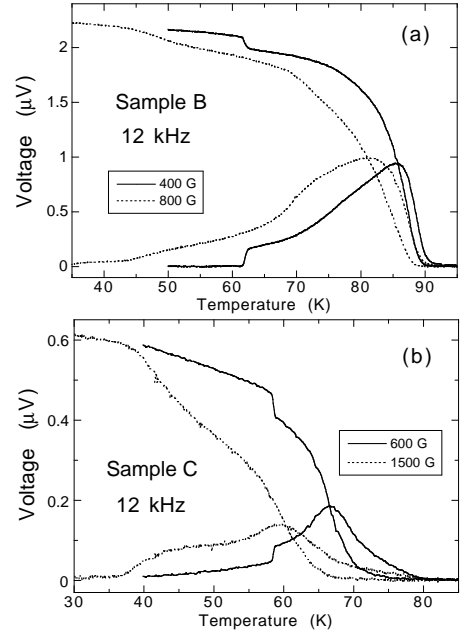}}
\vspace{2mm}
\caption{(a) and (b): In- and out-of-phase signals of samples B and C, respectively, in two selected magnetic fields below and above $H_{lim}$.}
\label{fig:4}
\end{figure}

After the existence of the first-order transition of the vortex system
in BSCCO has been established \cite{Zeldov}, 
much efforts have been devoted to 
the clarification of the details of the phase diagram.
There have been accumulating evidences that the FOT line is
a sublimation line, at which a solid of vortex lines transforms
into a gas of pancake vortices \cite{Matsuda,Fuchs-dc_flux}.
In the $H$-$T$ phase diagram, 
there are two lines other than the FOT line, called 
\lq \lq depinning line" and the \lq \lq second-peak line" \cite{Khaykovich}.
The three lines merge at the critical point; the depinning line 
separates the low- and high-temperature regions at fields above 
$H_{lim}$ and the second-peak line separates the high- and low-field 
regions at low temperatures.
Apparently, our $T_{k1}(H)$ and $T_{k2}(H)$ lines are very 
similar to the depinning line; thus, an examination of the 
$T_{k1}(H)$ and $T_{k2}(H)$ lines is expected to give an insight
into the nature of the depinning line.

Since the step-like change at $T_{d}(H)$ marks an abrupt onset of the 
long-range correlation in the vortex system, the broadened change
between $T_{k1}(H)$ and $T_{k2}(H)$ is expected to indicate an 
increase of a (short-range) correlation in the vortex system.
In general, a probe with higher frequency looks at physics at shorter
length scale \cite{Fisher}; in the case of our local ac response,
$\lambda_{ac}(\omega)$ is smaller for larger $\omega$.
With decreasing temperature, it is expected that the local ac response 
shows a qualitative change 
when the $c$-axis correlation length $L_c$ of the
vortex system starts to grow, and another qualitative change at a 
lower temperature is also expected when $L_c$ becomes comparable to 
$\lambda_{ac}(\omega)$.
This is one possible scenario for what is happening at $T_{k2}$ 
and $T_{k1}$.
The facts that $T_{k1}$ and $T_{k2}$ are dependent on frequency and that
a higher frequency gives a higher apparent $T_{k1}$ are
consistent with the above scenario.

\begin{figure}[t]
\epsfxsize=7.0cm
\centerline{\epsffile{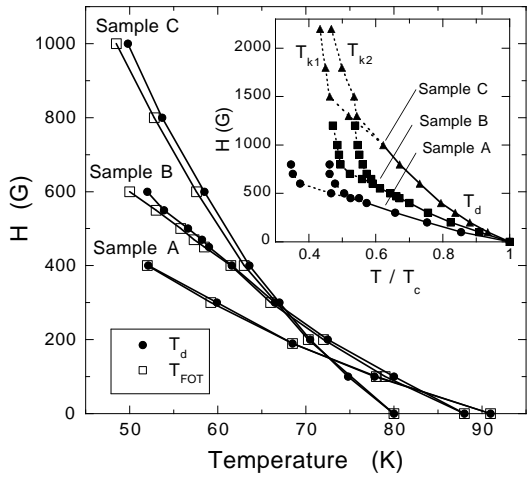}}
\vspace{2mm}
\caption{$T_{d}(H)$ lines and the $T_{FOT}(H)$ lines for the three samples.
Inset: $T_{d}(H)$, $T_{k1}(H)$, and $T_{k2}(H)$ lines 
(determined at 12 kHz), plotted vs $T/T_c$.}
\label{fig:5}
\end{figure}

Recently, Fuchs {\it et al.} used the change in the surface-barrier 
height for the determination of the vortex phase transformations 
\cite{Fuchs-Tx} (note here that the surface barrier is different from 
the geometrical barrier which is only effective at low fields near
$H_{c1}$) and the presence of a new transition line, $T_x$ line, 
at temperatures higher than the depinning line (and above the FOT line) 
was suggested.  
Since it is almost clear that the vortex phase above the FOT line is a 
gas of pancake vortices at temperatures higher than this new $T_x$ line 
\cite{Fuchs-Tx}, the existence of the $T_x$ line implies that 
the depinning line separates
a highly disordered entangled vortex solid (low-temperature side)
from either (a) disentangled liquid of lines with hexatic order 
or (b) some kind of solid which consists of an aligned stack of 
ordered two-dimensional pancake layers \cite{Fuchs-Tx}.
Our data suggests that the latter possibility (b) is more likely,
because the growth of the short-range correlation between 
$T_{k1}$ and $T_{k2}$ has a natural meaning of a growth of the
alignment of the pancake layers in the latter picture.
Note that we did not observe any feature which can be associated with  
the $T_x$ line;
this is reasonable because the $T_x$ line only manifests itself 
in a change in the surface barrier, which has little effect on 
our measurement.

Finally, let us briefly discuss the magnetic-field dependence of $T_{d}$.  
As has been reported \cite{Ando,Doyle}, the $T_{d}(H)$ line measured 
with the two-coil technique can be well fitted with the formula for 
the decoupling line \cite{Glazman,Daemen,Ikeda}. 
This is actually a matter of course, because our $T_{d}(H)$ line is 
identical to the $T_{FOT}$ line and the FOT is most likely to be 
a sublimation transition, which is essentially a decoupling transition.
Fittings of our data to the decoupling formula \cite{Glazman,Daemen}
$H \simeq H_0(T_c -T_d)/T_d$ give the anisotropy ratio $\gamma$
of $\sim$100, $\sim$85, and $\sim$77 for samples A, B, and C, respectively 
(the prefactor is given by 
$H_0\simeq \alpha_D \gamma^2 \phi_0^3/(4\pi \lambda(0))^2 T_c d$,
where $\alpha_D \simeq$0.1 is a constant, 
$d$=15 ${\rm \AA}$ is the spacing between the bilayers, 
and $\lambda(0) \simeq$2000 ${\rm \AA}$ is the penetration depth).

In summary, we measured the local ac response of three BSCCO crystals
(optimally doped, lightly overdoped, and overdoped samples)
using a miniature two-coil technique and compared the result with
a global dc magnetization measurement.
The origin of the step-like change in the two-coil measurement is 
identified to be the first-order transition
(FOT), where the {\it bulk} resistivity
(which is free from the edge contribution) is reported to show a 
sharp change \cite{Fuchs-bulk/strip}.
The sudden step-like change in the two-coil signal starts to be
broadened at fields above $H_{lim}$, where the FOT is no longer observed.
This broadened change takes place between $T_{k1}$ and $T_{k2}$ and
these two temperatures are still well defined, although they are
frequency dependent.
We discuss that the observation of the feature at $T_{k1}$ and $T_{k2}$ 
is likely to indicate the growth of a short-range correlation of the
vortex matter, which gives a clue to identify the nature of the depinning
line.

%
\medskip
\vfil

\end{document}